\documentclass[twocolumn]{revtex4}
\usepackage{mathbbold}
\usepackage{amsfonts}
\usepackage{amsmath}
\usepackage{amssymb}
\usepackage{charter}
\usepackage{graphicx}

\begin{document}

\title{Multiphoton states engineering by heralded interference via six-port
Mach-Zehnder interferometer}
\author{Qiang Ke$^{1}$, Xue-feng Zhan$^{1}$, Min-xiang Li$^{2}$, and
Xue-xiang Xu$^{1,\dag }$ }
\affiliation{$^{1}$College of Physics and Communication Electronics, Jiangxi Normal
University, Nanchang 330022, China;\\
$^{2}$School of Education, Jiangxi Normal University, Nanchang 330022, China%
\\
$^{\dag }$xuxuexiang@jxnu.edu.cn}

\begin{abstract}
Based on heralded interference on a six-port Mach-Zehnder interferometer, we
propose protocols to generate a series of multiphoton states in primary
output port, by injecting a coherent state in primary input port and two
Fock states in two ancillary input ports, and measuring two Fock states in
two ancillary output ports. Only manipulating at the single-photon level
(i.e, $\left\vert 0\right\rangle $ or $\left\vert 1\right\rangle $) in all
ancillary ports, we generate sixteen types (six categories) of multiphoton
nonclassical states, whose state vectors are unified as superposition of a
new coherent state, a single-photon added coherent state, and a two-photon
added coherent state. Indeed, a wide range of nonclassical phenomena can be
created by modulating the interaction parameters (including coherent field
strength and shift phase). We mainly analyze quadrature-squeezing effects
for all our considered states. Of particular interest is maximum squeezing
of up to 2.57\textrm{dB}, with success probability $6.7\%$, at least in our
present cases.
\end{abstract}

\maketitle

\section{Introduction}

Many quantum phenomena are based on the principle of quantum interference,
which lies at the heart of quantum theory\cite{1,2,3}. Quantum interference
has been widely applied in quantum technologies, such as quantum computation%
\cite{4}, quantum metrology\cite{5}, quantum teleporation\cite{6} and
quantum state engineering\cite{7}. In previous years, there are many typical
works on classical or quantum interference. The famous Yang's double-slit
experiment demonstrated the interference effect of light\cite{8}. The
landmark HOM experiment demonstrated the two-photon interference based on a
50:50 beamsplitter\cite{9}. The Knill-Laflamme-Milburn (KLM) scheme with
photon interference revealed that efficient quantum computation is possible
using only beam splitters, phase shifters, single-photon sources and
photo-detectors\cite{10}.

As we know, the beam splitter is a typical two-input-port splitter, which
can be generalized to multi-port beam splitters, such as tritter and quarter%
\cite{11}. Moreover, multi-port beam splitters are cornerstone devices for
high-dimensional quantum information tasks\cite{12}. Undeniably, these
devices have their respective unitary operations, which can be modeled and
implemented in the field of quantum optics\cite{13}. Indeed, any unitary
operation can be implemented by reconfigurable devices\cite{14,15,16}.
Following the construction of Mach-Zehnder interferometer (MZI), we can
extend the generalized Mach-Zehnder structure, which can be realized by a
chain of two subsequent multiport beam splitters. Thus, a multi-arm
interferometer can be realized by cascading two multi-port beam splitters.
Recently, Spagnolo et al. reported the experimental observation of
three-photon interference in an integrated three-port directional couplers%
\cite{17} and introduced three-dimensional (3D) multiphoton interferometry%
\cite{18}.

In the realm of quantum state engineering, many protocols are proposed to
prepare quantum states via quantum interference. For example, Bartley
developed a technique for generating multiphoton nonclassical states via
interference between coherent and Fock states\cite{19}. In one of our
previous works, we proposed a scheme to generate nonclassical states from a
coherent state heralded by KLM-type interference. The maximum squeezing of
our generated states is about 1.97928dB\cite{20}. In another work, based on
a conditional interferometry proposed by Paris\cite{21}, we prepare any
chosen superposition of the vacuum, one-photon, and two photon states\cite%
{22}. Recently, Ripala et al. presented a theoretical study of the
interferences between hybrid single-photon state, coherent state and vacuum
state based on a six-port MZI (6p-MZI)\cite{23}.

In this work, we also consider using another 6p-MZI as the interference
device and prepare quantum states by heralded method. The remaining paper is
organized as follows. In Sec.2, we introduce the 6p-MZI including its device
and transformation and then propose the generating protocol of multiphoton
quantum states. In Sec.3 we give the unified forms of state vectors, density
operators and success probabilities for all generated states considered in
this work. Then in sec.4, we analyze the quadrature-squeezing effects for
our considered quantum states. Conclusion is given in the last section.

\section{Generating protocols}

We consider a 6p-MZI as shown in Fig.1, which is composed of tritter 1,
phase shifter and tritter 2. Assuming these three devices can be described
by operators $\hat{T}_{1}$, $\hat{P}_{\phi }$, and $\hat{T}_{2}$,
respectively. Their respective transfer matrices $U_{\hat{T}_{2}}$, $U_{\hat{%
P}_{\phi }}$, and $U_{\hat{T}_{2}}$ are given in Appendix A, where the
tritters are symmetrical and lossless. Then, the 6p-MZI can be described by
operator $\hat{T}=\hat{T}_{2}\hat{P}_{\phi }\hat{T}_{1}$ and satisfy
\begin{equation}
\hat{T}\left(
\begin{array}{c}
a_{1}^{\dag } \\
a_{2}^{\dag } \\
a_{3}^{\dag }%
\end{array}%
\right) \hat{T}^{\dag }=U\left(
\begin{array}{c}
a_{1}^{\dag } \\
a_{2}^{\dag } \\
a_{3}^{\dag }%
\end{array}%
\right) ,  \label{1-1}
\end{equation}%
with transfer matrix%
\begin{equation}
U=U_{\hat{T}_{2}}U_{\hat{P}_{\phi }}U_{\hat{T}_{1}}=\left(
\begin{array}{ccc}
u_{11} & u_{12} & u_{13} \\
u_{21} & u_{22} & u_{23} \\
u_{31} & u_{32} & u_{33}%
\end{array}%
\right) ,  \label{1-2}
\end{equation}%
where $u_{11}=u_{22}=u_{33}=\left( e^{-i\phi }+2\right) /3$ and $%
u_{12}=u_{21}=u_{13}=u_{31}=u_{23}=u_{32}=\left( e^{-i\phi }-1\right) /3$.
\begin{figure}[tbp]
\label{Fig1} \centering\includegraphics[width=1.0\columnwidth]{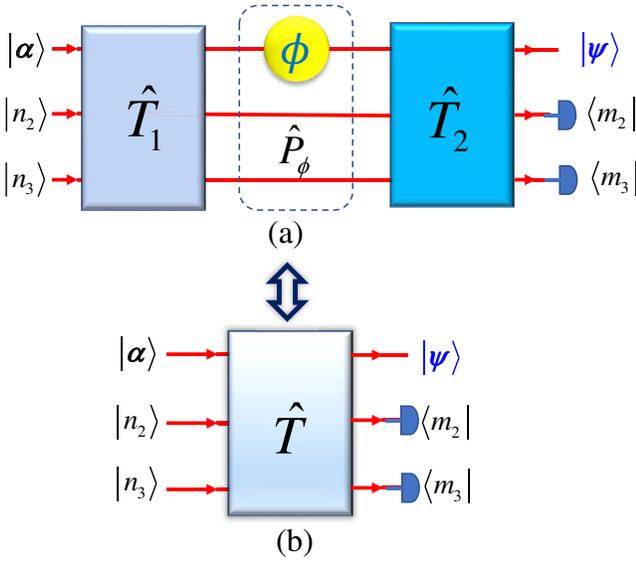}
\caption{(Colour online) (a) Schematic setup of 6p-MZI, which consist of
tritter 1 (by operator $\hat{T}_{1}$), phase shifter (by operator $\hat{P}_{%
\protect\phi }$) and tritter 2 (by operator $\hat{T}_{2}$); (b) Equivalent
of 6p-MZI (by operator $\hat{T}=\hat{T}_{2}\hat{P}_{\protect\phi }\hat{T}%
_{1} $). By injecting $\left\vert \protect\alpha \right\rangle $, $%
\left\vert n_{2}\right\rangle $, $\left\vert n_{3}\right\rangle $ in three
input ports, and measuring $\left\vert m_{2}\right\rangle $ and $\left\vert
m_{3}\right\rangle $ in two output ports, a new state $\left\vert \protect%
\psi \right\rangle $ will be generated in another output port. In contrast
with the input $\left\vert \protect\alpha \right\rangle $, the output $%
\left\vert \protect\psi \right\rangle $ will have a wide range of
nonclassical properties.}
\end{figure}

Based on the above 6p-MZI, we propose schemes to generate states according
to the following procedure:

(1) injecting coherent state $\left\vert \alpha \right\rangle $, Fock state $%
\left\vert n_{2}\right\rangle $ and Fock state $\left\vert
n_{3}\right\rangle $ in the first, second and third input port,
respectively; Here we assume $\alpha =\left\vert \alpha \right\vert
e^{i\theta }$ with $\theta =0$.

(2) interacting with interference in the 6p-MZI;

(3) making perfect counting measurement with Fock state $\left\vert
m_{2}\right\rangle $ and Fock state $\left\vert m_{3}\right\rangle $ in the
second and third output port, respectively;

(4) then generating a new quantum state $\left\vert \psi _{n_{2}n_{3},\alpha
,\phi }^{m_{2}m_{3}}\right\rangle =\left\vert \psi \right\rangle $ in the
first output port.

Formally, the generating state can be expressed as%
\begin{equation}
\left\vert \psi \right\rangle =\frac{1}{\sqrt{p_{n_{2}n_{3},\alpha ,\phi
}^{m_{2}m_{3}}}}\left\langle m_{3}\right\vert _{3}\left\langle
m_{2}\right\vert _{2}\hat{T}\left\vert \alpha \right\rangle _{1}\left\vert
n_{2}\right\rangle _{2}\left\vert n_{3}\right\rangle _{3}  \label{1-3}
\end{equation}%
with the success probability $p_{n_{2}n_{3},\alpha ,\phi }^{m_{2}m_{3}}$. By
tuning the parameters of the interactions, namely, the Fock-state numbers $%
n_{2}$, $n_{3}$, $m_{2}$, $m_{3}$, the shifted phase $\phi $, and
coherent-state amplitude $\alpha $, a series of states can be generated.
General analytical results for state vector, density operator, success
probability, and expectation value for $\left\vert \psi \right\rangle $ can
be obtained from results provided in Appendix B and Appendix C. It should be
noted that operator $a_{1}^{\dag }$ ($a_{1}$) is directly written as $%
a^{\dag }$ ($a$).

\section{Considered multiphoton states}

Zero-photon operation and single-photon operation are known to dramatically
alter the properties of certain quantum optical states\cite{24,25}. They
have potential applications in the context of quantum communications and
quantum information processing\cite{26}. Enlighten by these ideas, we only
consider by taking $n_{2}$, $n_{3}$, $m_{2}$, $m_{3}$ $\in $ ($0$, $1$) in
our above protocol and study 16 types of generating states in this work.
After our careful analysis, we find that their state vectors, density
operators and success probabilities can be unified.

\textit{Unified state vectors:} State vectors of all considered generating
states can be unified as%
\begin{equation}
\left\vert \psi _{i}\right\rangle =\frac{1}{\sqrt{N_{i}}}(c_{i0}+c_{i1}a^{%
\dag }+c_{i2}a^{\dag 2})\left\vert u_{11}\alpha \right\rangle ,  \label{2-1}
\end{equation}%
with normalization factor%
\begin{eqnarray}
N_{i} &=&\left\vert c_{i0}\right\vert ^{2}+\left\vert c_{i1}\right\vert
^{2}+2\left\vert c_{i2}\right\vert ^{2}  \notag \\
&&+(\left\vert c_{i1}\right\vert ^{2}+4\left\vert c_{i2}\right\vert
^{2})\left\vert u_{11}\right\vert ^{2}\left\vert \alpha \right\vert ^{2}
\notag \\
&&+\left\vert c_{i2}\right\vert ^{2}\left\vert u_{11}\right\vert
^{4}\left\vert \alpha \right\vert ^{4}+2\text{Re}(c_{i0}c_{i2}^{\ast
}u_{11}^{2}\alpha ^{2})  \notag \\
&&+2\text{Re}[(c_{i0}c_{i1}^{\ast }+2c_{i1}c_{i2}^{\ast })u_{11}\alpha ]
\notag \\
&&+2\text{Re}(c_{i1}c_{i2}^{\ast }\left\vert u_{11}\right\vert
^{2}\left\vert \alpha \right\vert ^{2}u_{11}\alpha ).  \label{2-2}
\end{eqnarray}%
Here, $c_{i0}$, $c_{i1}$, and $c_{i2}$ for each $\left\vert \psi
_{i}\right\rangle $ are illustrated in Table I. These 16 states can\ be
classified into the following 6 categories:

(1) $\left\vert \psi _{1}\right\rangle $, $\left\vert \psi _{2}\right\rangle
$, $\left\vert \psi _{3}\right\rangle $, and $\left\vert \psi
_{4}\right\rangle $ are new CSs $\left\vert u_{11}\alpha \right\rangle $
(CS);

(2) $\left\vert \psi _{5}\right\rangle $ and $\left\vert \psi
_{6}\right\rangle $ are single-photon added CSs $a^{\dag }\left\vert
u_{11}\alpha \right\rangle $ (SPACS);

(3) $\left\vert \psi _{7}\right\rangle $\ is a two-photon added CS $a^{\dag
2}\left\vert u_{11}\alpha \right\rangle $ (TPACS);

(4) $\left\vert \psi _{8}\right\rangle $, $\left\vert \psi _{9}\right\rangle
$, $\left\vert \psi _{10}\right\rangle $, $\left\vert \psi
_{11}\right\rangle $, $\left\vert \psi _{12}\right\rangle $, and $\left\vert
\psi _{13}\right\rangle $ are superposition states of $\left\vert
u_{11}\alpha \right\rangle $ and $a^{\dag }\left\vert u_{11}\alpha
\right\rangle $;

(5) $\left\vert \psi _{14}\right\rangle $ and $\left\vert \psi
_{15}\right\rangle $ are superposition states of $a^{\dag }\left\vert
u_{11}\alpha \right\rangle $ and $a^{\dag 2}\left\vert u_{11}\alpha
\right\rangle $;

(6) $\left\vert \psi _{16}\right\rangle $ is a superposition state of $%
\left\vert u_{11}\alpha \right\rangle $, $a^{\dag }\left\vert u_{11}\alpha
\right\rangle $ and $a^{\dag 2}\left\vert u_{11}\alpha \right\rangle $.
\begin{table}[h]
\caption{$c_{i0}$, $c_{i1}$, and $c_{i2}$ for each $\left\vert \protect\psi %
_{i}\right\rangle $. Note that $\protect\tau _{1}=u_{12}u_{23}$ $%
+u_{13}u_{22}$, $\protect\tau _{2}=u_{12}u_{33}$ $+u_{13}u_{32}$, $\protect%
\tau _{3}=u_{21}u_{32}$ $+u_{22}u_{31}$, $\protect\tau _{4}=u_{21}u_{33}$ $%
+u_{23}u_{31}$, $\protect\tau _{5}=u_{23}u_{32}$ $+u_{22}u_{33}$, $\protect%
\kappa =u_{12}u_{23}u_{31}$ $+u_{13}u_{22}u_{31}$ $+u_{13}u_{21}u_{32}$ $%
+u_{12}u_{21}u_{33}$.}
\begin{center}
\begin{tabular}{|c|c|c|c|c|}
\hline\hline
$\left\vert \psi _{n_{2}n_{3},\alpha ,\phi }^{m_{2}m_{3}}\right\rangle $ & $%
\left\vert \psi _{i}\right\rangle $ & $c_{i0}$ & $c_{i1}$ & $c_{i2}$ \\
\hline\hline
$\left\vert \psi _{00,\alpha ,\phi }^{00}\right\rangle $ & $\left\vert \psi
_{1}\right\rangle $ & $1$ & $0$ & $0$ \\ \hline
$\left\vert \psi _{00,\alpha ,\phi }^{10}\right\rangle $ & $\left\vert \psi
_{2}\right\rangle $ & $u_{12}\alpha $ & $0$ & $0$ \\ \hline
$\left\vert \psi _{00,\alpha ,\phi }^{01}\right\rangle $ & $\left\vert \psi
_{3}\right\rangle $ & $u_{13}\alpha $ & $0$ & $0$ \\ \hline
$\left\vert \psi _{00,\alpha ,\phi }^{11}\right\rangle $ & $\left\vert \psi
_{4}\right\rangle $ & $u_{12}u_{13}\alpha ^{2}$ & $0$ & $0$ \\ \hline
$\left\vert \psi _{10,\alpha ,\phi }^{00}\right\rangle $ & $\left\vert \psi
_{5}\right\rangle $ & $0$ & $u_{21}$ & $0$ \\ \hline
$\left\vert \psi _{01,\alpha ,\phi }^{00}\right\rangle $ & $\left\vert \psi
_{6}\right\rangle $ & $0$ & $u_{31}$ & $0$ \\ \hline
$\left\vert \psi _{11,\alpha ,\phi }^{00}\right\rangle $ & $\left\vert \psi
_{7}\right\rangle $ & $0$ & $0$ & $u_{21}u_{31}$ \\ \hline
$\left\vert \psi _{10,\alpha ,\phi }^{10}\right\rangle $ & $\left\vert \psi
_{8}\right\rangle $ & $u_{22}$ & $u_{12}u_{21}\alpha $ & $0$ \\ \hline
$\left\vert \psi _{01,\alpha ,\phi }^{01}\right\rangle $ & $\left\vert \psi
_{9}\right\rangle $ & $u_{33}$ & $u_{13}u_{31}\alpha $ & $0$ \\ \hline
$\left\vert \psi _{01,\alpha ,\phi }^{10}\right\rangle $ & $\left\vert \psi
_{10}\right\rangle $ & $u_{32}$ & $u_{12}u_{31}\alpha $ & $0$ \\ \hline
$\left\vert \psi _{10,\alpha ,\phi }^{01}\right\rangle $ & $\left\vert \psi
_{11}\right\rangle $ & $u_{23}$ & $u_{13}u_{21}\alpha $ & $0$ \\ \hline
$\left\vert \psi _{10,\alpha ,\phi }^{11}\right\rangle $ & $\left\vert \psi
_{12}\right\rangle $ & $\tau _{1}\alpha $ & $u_{12}u_{13}u_{21}\alpha ^{2}$
& $0$ \\ \hline
$\left\vert \psi _{01,\alpha ,\phi }^{11}\right\rangle $ & $\left\vert \psi
_{13}\right\rangle $ & $\tau _{2}\alpha $ & $u_{12}u_{13}u_{31}\alpha ^{2}$
& $0$ \\ \hline
$\left\vert \psi _{11,\alpha ,\phi }^{10}\right\rangle $ & $\left\vert \psi
_{14}\right\rangle $ & $0$ & $\tau _{3}$ & $u_{12}u_{21}u_{31}\alpha $ \\
\hline
$\left\vert \psi _{11,\alpha ,\phi }^{01}\right\rangle $ & $\left\vert \psi
_{15}\right\rangle $ & $0$ & $\tau _{4}$ & $u_{13}u_{21}u_{31}\alpha $ \\
\hline
$\left\vert \psi _{11,\alpha ,\phi }^{11}\right\rangle $ & $\left\vert \psi
_{16}\right\rangle $ & $\tau _{5}$ & $\kappa \alpha $ & $%
u_{12}u_{13}u_{21}u_{31}\alpha ^{2}$ \\ \hline
\end{tabular}%
\end{center}
\end{table}

\textit{Unified density operators:} Density operators $\rho _{j}=\left\vert
\psi _{j}\right\rangle \left\langle \psi _{j}\right\vert $\ can be unified
as
\begin{eqnarray}
\rho _{i} &=&\frac{\left\vert c_{i0}\right\vert ^{2}}{N_{i}}\rho ^{(0,0)}+%
\frac{\left\vert c_{i1}\right\vert ^{2}}{N_{i}}\rho ^{(1,1)}+\frac{%
\left\vert c_{i2}\right\vert ^{2}}{N_{i}}\rho ^{(2,2)}  \notag \\
&&+\frac{c_{i0}c_{i1}^{\ast }}{N_{i}}\rho ^{(0,1)}+\frac{c_{i1}c_{i0}^{\ast }%
}{N_{i}}\rho ^{(1,0)}  \notag \\
&&+\frac{c_{i0}c_{i2}^{\ast }}{N_{i}}\rho ^{(0,2)}+\frac{c_{i2}c_{i0}^{\ast }%
}{N_{i}}\rho ^{(2,0)}  \notag \\
&&+\frac{c_{i1}c_{i2}^{\ast }}{N_{i}}\rho ^{(1,2)}+\frac{c_{i2}c_{i1}^{\ast }%
}{N_{i}}\rho ^{(2,1)}.  \label{2-3}
\end{eqnarray}%
for above 16 states after setting%
\begin{equation}
\rho ^{(h_{l},h_{r})}=a^{\dag h_{l}}\left\vert u_{11}\alpha \right\rangle
\left\langle u_{11}\alpha \right\vert a^{h_{r}},  \label{2-4}
\end{equation}%
where $h_{l}$ and $h_{r}$\ are non-negative integers. Thus, any property of $%
\rho _{i}$\ can be obtained from that of $\rho ^{(h_{l},h_{r})}$ according
to Eq.(\ref{2-3}).

\textit{Unified success probabilities:} Success probabilities of above
generated states can be unified as%
\begin{equation}
p_{n_{2}n_{3},\alpha ,\phi }^{m_{2}m_{3}}\longmapsto
p_{i}=N_{i}e^{(\left\vert u_{11}\right\vert ^{2}-1)\left\vert \alpha
\right\vert ^{2}}.  \label{2-5}
\end{equation}%
In Fig.2 and Fig3, we plot all $p_{i}$s in the ($\left\vert \alpha
\right\vert ,\phi $) space with $\phi \in \lbrack 0,\pi ]$, where different
colors denote different categories. We find that even for the states in the
same category, the success probability may be different. For example, in the
first category $\left\vert u_{11}\alpha \right\rangle $ including four
states $\left\vert \psi _{1}\right\rangle $, $\left\vert \psi
_{2}\right\rangle $, $\left\vert \psi _{3}\right\rangle $, and $\left\vert
\psi _{4}\right\rangle $, there are three different success probabilities,
i.e., $p_{1}$, $p_{2}=p_{3}$, and $p_{4}$. In order to see clearly the
curves in above three-dimensional plots, we take $\left\vert \alpha
\right\vert =2$ as example and plot each $p_{i}$\ as a function of $\phi $
in Fig.4. In the interval of $[0,2\pi ]$, the probabilities are symmetrical
on $\phi =\pi $.
\begin{figure}[tbp]
\label{Fig2} \centering\includegraphics[width=1.0\columnwidth]{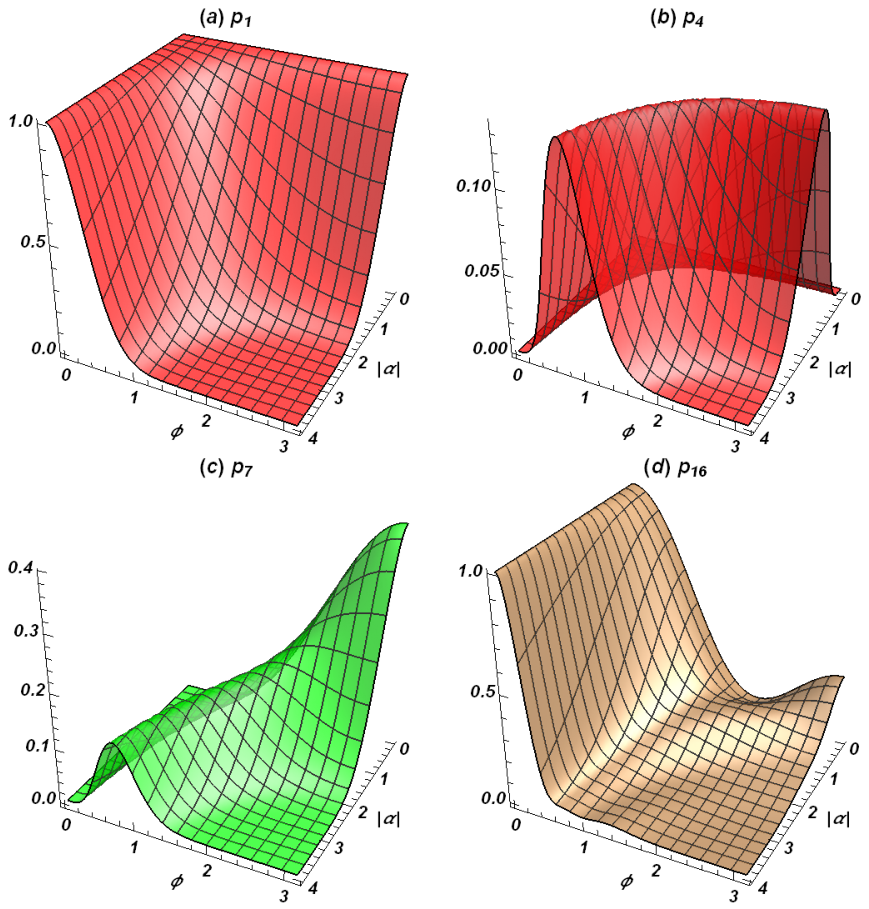}
\caption{Success probabilities in ($\left\vert \protect\alpha \right\vert $,
$\protect\phi $) space for states: (a) $\left\vert \protect\psi %
_{1}\right\rangle $; (b) $\left\vert \protect\psi _{4}\right\rangle $; (c) $%
\left\vert \protect\psi _{7}\right\rangle $; (d) $\left\vert \protect\psi %
_{16}\right\rangle $.}
\end{figure}
\begin{figure}[tbp]
\label{Fig3} \centering\includegraphics[width=1.0\columnwidth]{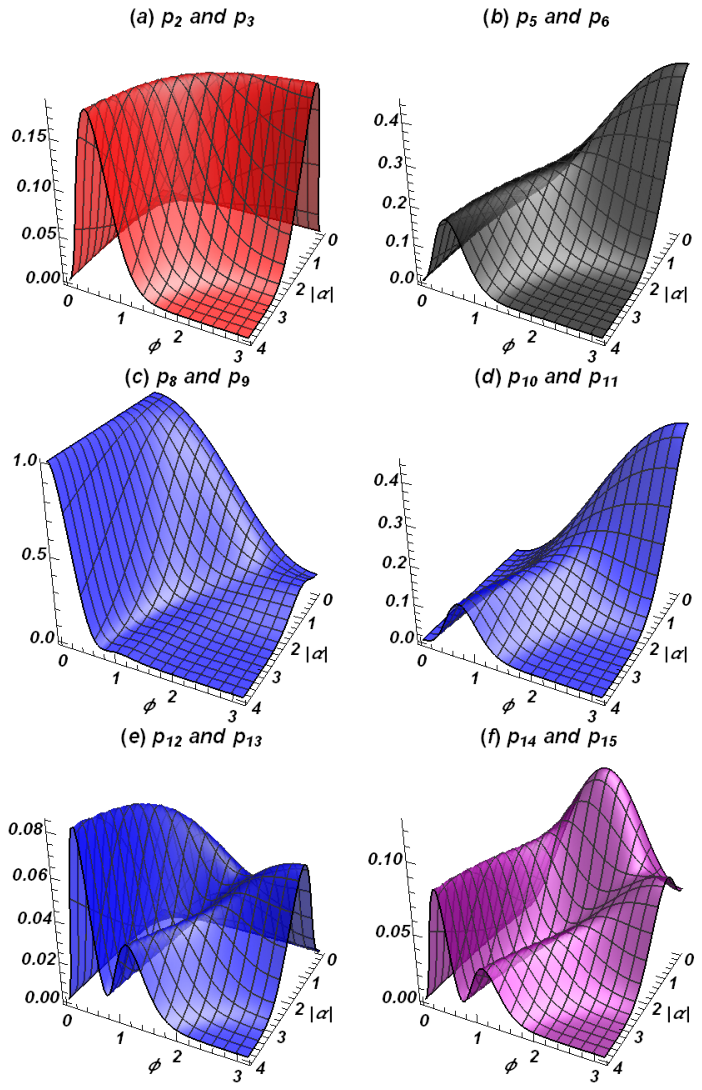}
\caption{Success probabilities in ($\left\vert \protect\alpha \right\vert $,
$\protect\phi $) space for states: (a) $\left\vert \protect\psi %
_{2,3}\right\rangle $; (b) $\left\vert \protect\psi _{5,6}\right\rangle $;
(c) $\left\vert \protect\psi _{8,9}\right\rangle $; (d) $\left\vert \protect%
\psi _{10,11}\right\rangle $; (e) $\left\vert \protect\psi %
_{12,13}\right\rangle $; (f) $\left\vert \protect\psi _{14,15}\right\rangle $%
.}
\end{figure}
\begin{figure}[tbp]
\label{Fig4} \centering\includegraphics[width=1.0\columnwidth]{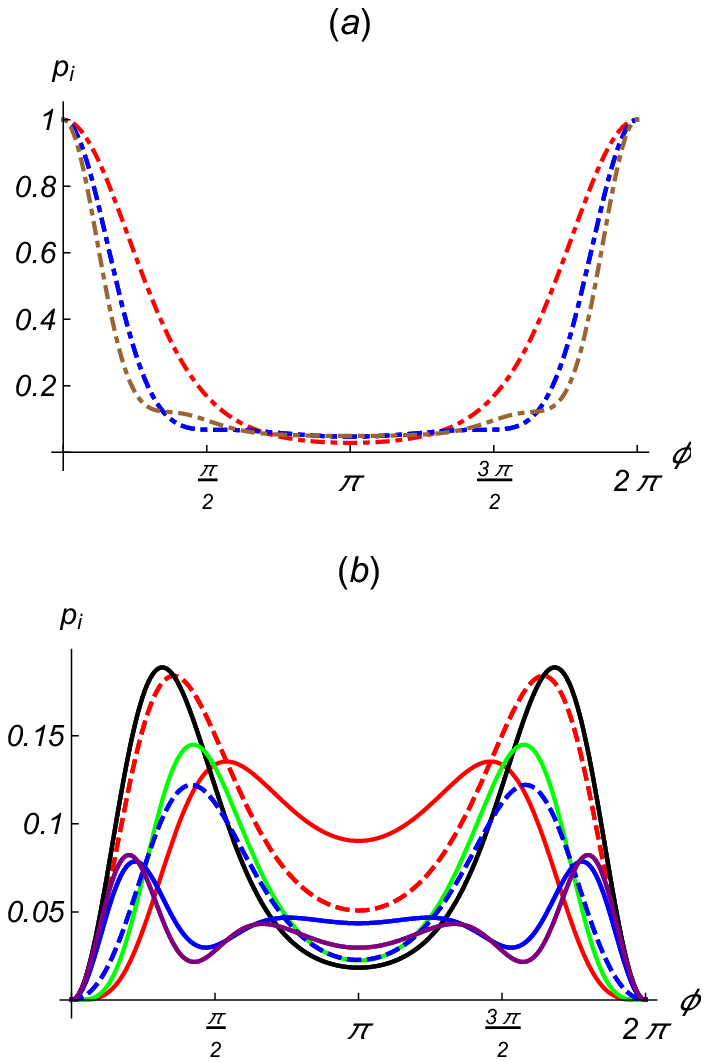}
\caption{(a) $p_{i}$ as a function of $\protect\phi $, with $\left\vert
\protect\alpha \right\vert =2$ for $\left\vert \protect\psi %
_{1}\right\rangle $ (red dotdashed), $\left\vert \protect\psi %
_{8,9}\right\rangle $ (blue dotdashed), and $\left\vert \protect\psi %
_{16}\right\rangle $ (brown dotdashed); (b) $p_{i}$ as a function of $%
\protect\phi $, with $\left\vert \protect\alpha \right\vert =2$ for $%
\left\vert \protect\psi _{2,3}\right\rangle $ (red dashed), $\left\vert
\protect\psi _{4}\right\rangle $ (red), $\left\vert \protect\psi %
_{5,6}\right\rangle $ (black), $\left\vert \protect\psi _{7}\right\rangle $
(green), $\left\vert \protect\psi _{10,11}\right\rangle $ (blue dashed), $%
\left\vert \protect\psi _{12,13}\right\rangle $ (blue), and $\left\vert
\protect\psi _{14,15}\right\rangle $ (purple).}
\end{figure}

\section{Quadrature-squeezing effects}

Quadrature-squeezing effects can be judged from that the $\hat{x}$%
-quadrature\ variance $\left\langle (\Delta \hat{x})^{2}\right\rangle $\
below the standard limit 0.5, where $\hat{x}=(a+a^{\dag })/\sqrt{2}$. That
is, a state is quadrature-squeezing if $\left\langle (\Delta \hat{x}%
)^{2}\right\rangle <0.5$\cite{27}. Obviously, $\left\langle (\Delta \hat{x}%
)^{2}\right\rangle =0.5$ is always right for $\left\vert \psi
_{1}\right\rangle $, $\left\vert \psi _{2}\right\rangle $, $\left\vert \psi
_{3}\right\rangle $, and $\left\vert \psi _{4}\right\rangle $. But for
states from $\left\vert \psi _{5}\right\rangle $\ to $\left\vert \psi
_{16}\right\rangle $, are they quadrature-squeezing? Fig.5 and Fig.6(a)
presents their feasibility squeezing regions\ in the ($\left\vert \alpha
\right\vert $, $\phi $) space. Further, as examples, we plot $\left\langle
(\Delta \hat{x})^{2}\right\rangle $s\ of $\left\vert \psi _{16}\right\rangle
$ with $\left\vert \alpha \right\vert =2,4,6$ as functions of $\phi $ in
Fig.6(b).
\begin{figure}[tbp]
\label{Figsq1} \centering\includegraphics[width=1.0\columnwidth]{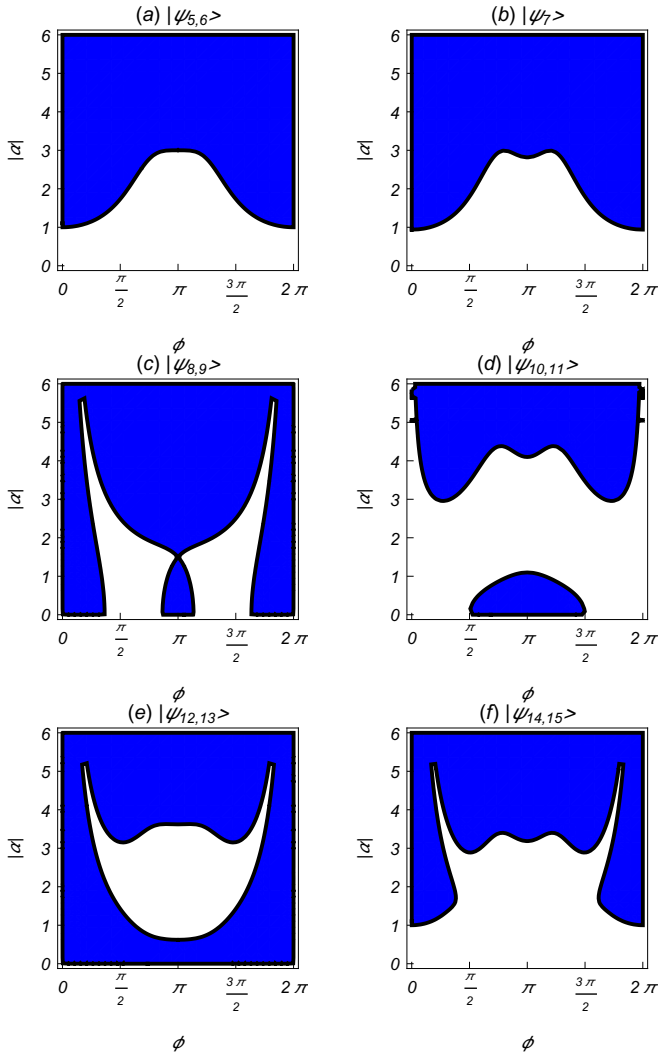}
\caption{Feasibility $\hat{x}$-squeezing effects in ($\left\vert \protect%
\alpha \right\vert $, $\protect\phi $) space for states from $\left\vert
\protect\psi _{5}\right\rangle $ to $\left\vert \protect\psi %
_{15}\right\rangle $, showing by blue regions.}
\end{figure}
\begin{figure}[tbp]
\label{Figsq2} \centering\includegraphics[width=1.0\columnwidth]{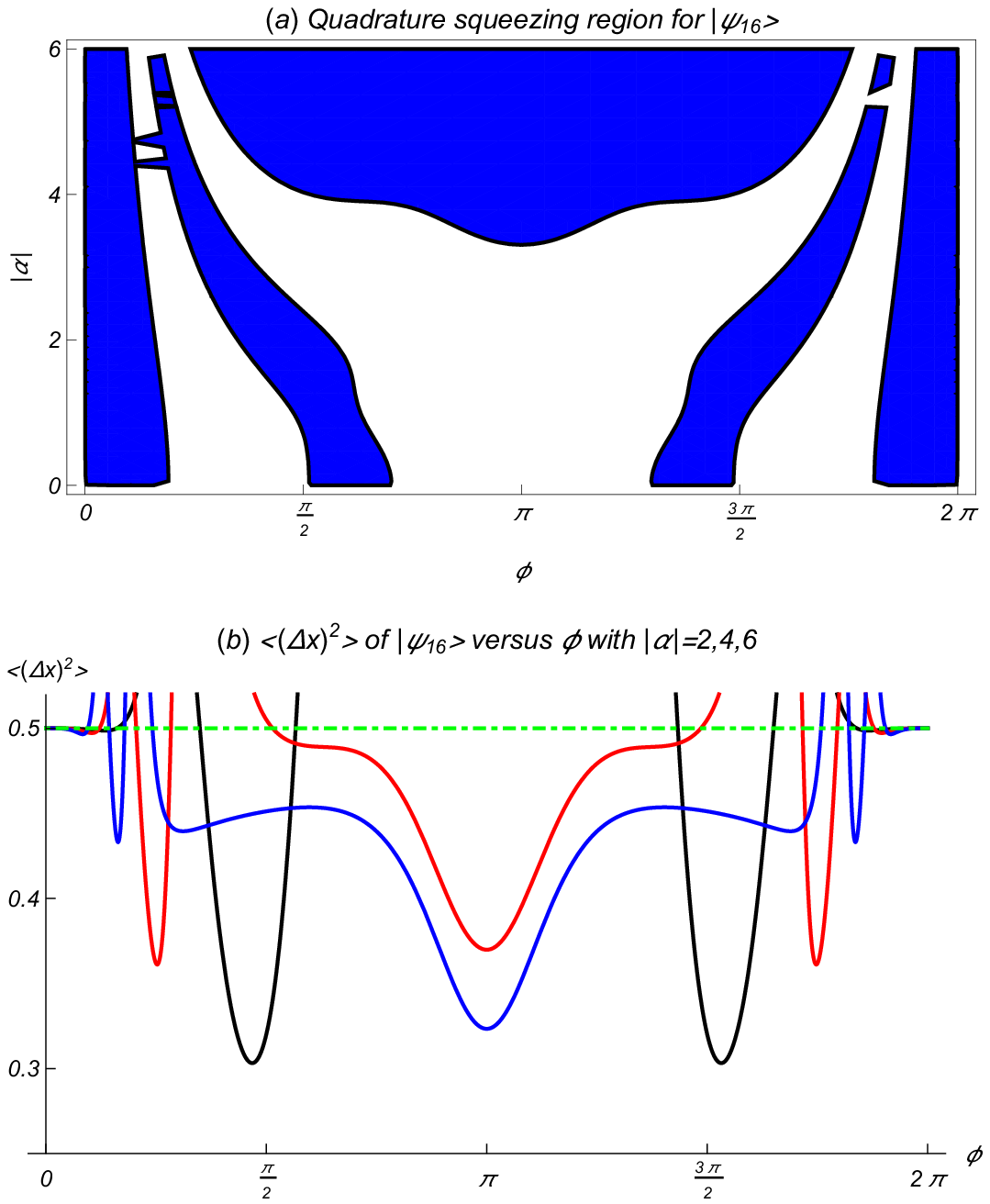}
\caption{(a) Feasibility $\hat{x}$-squeezing effects in ($\left\vert \protect%
\alpha \right\vert $, $\protect\phi $) space for state $\left\vert \protect%
\psi _{16}\right\rangle $, see blue region; (b) $\left\langle (\Delta \hat{x}%
)^{2}\right\rangle $ as a function of $\protect\phi $ for state $\left\vert
\protect\psi _{16}\right\rangle $\ with $\left\vert \protect\alpha %
\right\vert =2,4,6$.}
\end{figure}

The quadrature variance $\left\langle (\Delta \hat{x})^{2}\right\rangle $ of
a state, relative to the vacuum noise level of $0.5$, can show squeezing
degree by a logarithmic scale (units of $\mathrm{dB}$) through $-10\log
_{10}\left( \left\langle (\Delta \hat{x})^{2}\right\rangle /0.5\right) $. In
order to see the maximum squeezing degrees for states from $\left\vert \psi
_{5}\right\rangle $\ to $\left\vert \psi _{16}\right\rangle $, we use the
computing software (MATHEMATICA) to find the minimize $\left\langle (\Delta
\hat{x})^{2}\right\rangle $ values under\ the constraints $0\leq \left\vert
\alpha \right\vert \leq 10$\ and $0\leq \phi \leq 2\pi $. The numerical or
analytical results show that:

(1) For states $\left\vert \psi _{1}\right\rangle $, $\left\vert \psi
_{2}\right\rangle $, $\left\vert \psi _{3}\right\rangle $, and $\left\vert
\psi _{4}\right\rangle $, the squeezing degrees are $0$\textrm{dB}, showing
without squeezing, because they are all coherent states;

(2) For states including $\left\vert \psi _{5}\right\rangle $, $\left\vert
\psi _{6}\right\rangle $, $\left\vert \psi _{8}\right\rangle $, $\left\vert
\psi _{9}\right\rangle $, $\left\vert \psi _{10}\right\rangle $, $\left\vert
\psi _{11}\right\rangle $, $\left\vert \psi _{12}\right\rangle $, and $%
\left\vert \psi _{13}\right\rangle $, their minimum variances ($\sim 0.375$)
correspond to $1.25$\textrm{dB} squeezing;

(3) For states including $\left\vert \psi _{7}\right\rangle $, $\left\vert
\psi _{14}\right\rangle $, and $\left\vert \psi _{15}\right\rangle $, their
minimum variances ($\sim 0.322$) correspond to about $1.91$\textrm{dB}
squeezing;

(4) For state $\left\vert \psi _{16}\right\rangle $, its minimum variance ($%
\sim 0.277$) corresponds to $2.57$\textrm{dB} squeezing.\textrm{\ }

Interestingly, squeezing degrees with $1.91$\textrm{dB}\ and $2.57$\textrm{dB%
}\ are meaningful results in our proposals. Compared with the $1.25$\textrm{%
dB}\ squeezing degree in the Bartley's work\cite{19}, these maximum
squeezing degrees have been greatly improved. Moreover, the success
probability to obtain $2.57$\textrm{dB}\ squeezing can reach $6.7\%$.

Since the $\hat{p}$-quadratures are never squeezed for each $\left\vert \psi
_{i}\right\rangle $, we don't discuss $\hat{p}$-quadrature\ variance $%
\left\langle (\Delta \hat{p})^{2}\right\rangle $, with $\hat{p}=(a-a^{\dag
})/(i\sqrt{2})$.

\section{Conclusion}

16 types of multiphoton states have been produced by available schemes based
on a given 6p-MZI. Of course, these 16 states can be classified into 6
categories, according to their components of CS, SPACS and TPACS. Moreover,
we analytically unified their state vectors, density operators, success
probabilities. We numerically analyzed $\hat{x}$-squeezing effects and found
that a maximum squeezing degree with $2.57$\textrm{dB} can be obtained. In
fact, many other interesting phenomena can be exhibited by taking different
parameters. In view of the complexity and difficulty of calculation, we
don't consider cases with bigger $n_{2}$, $n_{3}$, $m_{2}$, $m_{3}$ in this
work. However, we don't considered arbitrary devices with asymmetrical
tritters.

Rather than application in quantum state engineering, these multiport
splitters can also be applied in quantum simulations\cite{28}, linear
optical computing\cite{29}\ and nonlocality tests\cite{30}. They will be
important tools in quantum fields. We believe that the subject of multiport
splitters will become a very active area of research.

\subsection*{\textbf{Appendix A: Devices and transformations}}

We introduce devices and transformations used in our present scheme.

\textit{(1) Tritter 1:} Transformation of Tritter 1 is described by%
\begin{equation}
\hat{T}_{1}\left(
\begin{array}{c}
a_{1}^{\dag } \\
a_{2}^{\dag } \\
a_{3}^{\dag }%
\end{array}%
\right) \hat{T}_{1}^{\dag }=U_{\hat{T}_{1}}\left(
\begin{array}{c}
a_{1}^{\dag } \\
a_{2}^{\dag } \\
a_{3}^{\dag }%
\end{array}%
\right) ,  \tag{A.1}
\end{equation}%
with transfer matrix%
\begin{equation}
U_{\hat{T}_{1}}=\frac{1}{\sqrt{3}}\left(
\begin{array}{ccc}
1 & 1 & 1 \\
1 & e^{i\frac{2\pi }{3}} & e^{i\frac{4\pi }{3}} \\
1 & e^{i\frac{4\pi }{3}} & e^{i\frac{2\pi }{3}}%
\end{array}%
\right)  \tag{A.2}
\end{equation}

\textit{(2) Phase shifter:} Transformation of phase shifter is described by%
\begin{equation}
\hat{P}_{\phi }\left(
\begin{array}{c}
a_{1}^{\dag } \\
a_{2}^{\dag } \\
a_{3}^{\dag }%
\end{array}%
\right) \hat{P}_{\phi }^{\dag }=U_{\hat{P}_{\phi }}\left(
\begin{array}{c}
a_{1}^{\dag } \\
a_{2}^{\dag } \\
a_{3}^{\dag }%
\end{array}%
\right) ,  \tag{A.3}
\end{equation}%
with transfer matrix
\begin{equation}
U_{\hat{P}_{\phi }}=\left(
\begin{array}{ccc}
e^{-i\phi } & 0 & 0 \\
0 & 1 & 0 \\
0 & 0 & 1%
\end{array}%
\right)  \tag{A.4}
\end{equation}

\textit{(3) Tritter 2:} Transformation of Tritter 2 is described by%
\begin{equation}
\hat{T}_{2}\left(
\begin{array}{c}
a_{1}^{\dag } \\
a_{2}^{\dag } \\
a_{3}^{\dag }%
\end{array}%
\right) \hat{T}_{2}^{\dag }=U_{\hat{T}_{2}}\left(
\begin{array}{c}
a_{1}^{\dag } \\
a_{2}^{\dag } \\
a_{3}^{\dag }%
\end{array}%
\right) ,  \tag{A.5}
\end{equation}%
with transfer matrix
\begin{equation}
U_{\hat{T}_{2}}=\frac{1}{\sqrt{3}}\left(
\begin{array}{ccc}
1 & 1 & 1 \\
1 & e^{i\frac{4\pi }{3}} & e^{i\frac{2\pi }{3}} \\
1 & e^{i\frac{2\pi }{3}} & e^{i\frac{4\pi }{3}}%
\end{array}%
\right)  \tag{A.6}
\end{equation}

\subsection*{\textbf{Appendix B: About generated states}}

In this appendix, we give the general expression of state vector, density
operator and success probability for all generated states.

\textit{(1) State vector:}\textbf{\ }Noting $\left\vert \alpha \right\rangle
=e^{-\left\vert \alpha \right\vert ^{2}/2}e^{\alpha a_{1}^{\dag }}\left\vert
0\right\rangle _{1}$, $\left\vert n_{j}\right\rangle =$ $\left(
n_{j}!\right) ^{-1/2}$ $\partial _{s_{j}}^{n_{j}}e^{s_{j}a_{j}^{\dag
}}\left\vert 0\right\rangle _{j}|_{s_{j}=0}$, and $\left\langle
m_{j}\right\vert =$ $\left( m_{j}!\right) ^{-1/2}$ $\partial
_{t_{j}}^{m_{j}}\left\langle 0\right\vert _{j}e^{t_{j}a_{j}}|_{t_{j}=0}$,
and using the transformation in Eq.(\ref{1-1}) and $\hat{T}\left\vert
0\right\rangle _{1}\left\vert 0\right\rangle _{2}\left\vert 0\right\rangle
_{3}$ $=\left\vert 0\right\rangle _{1}\left\vert 0\right\rangle
_{2}\left\vert 0\right\rangle _{3}$, we obtain

\begin{subequations}
\begin{align}
\left\vert \psi \right\rangle & =\frac{e^{-\frac{\left\vert \alpha
\right\vert ^{2}}{2}}}{\sqrt{m_{3}!m_{2}!n_{2}!n_{3}!p_{n_{2}n_{3},\alpha
,\phi }^{m_{2}m_{3}}}}\partial _{s_{2}}^{n_{2}}\partial
_{s_{3}}^{n_{3}}\partial _{t_{2}}^{m_{2}}\partial _{t_{3}}^{m_{3}}  \notag \\
& e^{t_{2}\left( \alpha u_{12}+s_{2}u_{22}+s_{3}u_{32}\right) +t_{3}\left(
\alpha u_{13}+s_{2}u_{23}+\allowbreak s_{3}u_{33}\right) }  \notag \\
& e^{\left( \alpha u_{11}+s_{2}u_{21}+\allowbreak s_{3}u_{31}\right)
a_{1}^{\dag }}\left\vert 0\right\rangle _{1}|_{(s_{2},t_{2},s_{3},t_{3})=0}
\tag{B.1}
\end{align}%
whose conjugate state is
\end{subequations}
\begin{subequations}
\begin{align}
\left\langle \psi \right\vert & =\frac{e^{-\frac{\left\vert \alpha
\right\vert ^{2}}{2}}}{\sqrt{m_{3}!m_{2}!n_{2}!n_{3}!p_{n_{2}n_{3},\alpha
,\phi }^{m_{2}m_{3}}}}\partial _{f_{2}}^{n_{2}}\partial
_{f_{3}}^{n_{3}}\partial _{g_{2}}^{m_{2}}\partial _{g_{3}}^{m_{3}}  \notag \\
& e^{g_{2}\left( \alpha ^{\ast }u_{12}^{\ast }+f_{2}u_{22}^{\ast
}+f_{3}u_{32}^{\ast }\right) +g_{3}\left( \alpha ^{\ast }u_{13}^{\ast
}+f_{2}u_{23}^{\ast }+\allowbreak f_{3}u_{33}^{\ast }\right) }  \notag \\
& \left\langle 0\right\vert _{1}e^{\left( \alpha ^{\ast }u_{11}^{\ast
}+f_{2}u_{21}^{\ast }+\allowbreak f_{3}u_{31}^{\ast }\right)
a_{1}}|_{(f_{2},g_{2},f_{3},g_{3})=0}.  \tag{B.2}
\end{align}

\textit{(2) Density operator:}\textbf{\ }The general density operator $\rho
=\left\vert \psi \right\rangle \left\langle \psi \right\vert $ can be
written as
\end{subequations}
\begin{subequations}
\begin{align}
\rho & =\frac{e^{-\left\vert \alpha \right\vert ^{2}}}{%
m_{3}!m_{2}!n_{2}!n_{3}!p_{n_{2}n_{3},\alpha ,\phi }^{m_{2}m_{3}}}  \notag \\
& \partial _{s_{2}}^{n_{2}}\partial _{s_{3}}^{n_{3}}\partial
_{t_{2}}^{m_{2}}\partial _{t_{3}}^{m_{3}}\partial _{f_{2}}^{n_{2}}\partial
_{f_{3}}^{n_{3}}\partial _{g_{2}}^{m_{2}}\partial _{g_{3}}^{m_{3}}  \notag \\
& e^{t_{2}\left( \alpha u_{12}+s_{2}u_{22}+s_{3}u_{32}\right) +t_{3}\left(
\alpha u_{13}+s_{2}u_{23}+\allowbreak s_{3}u_{33}\right) }  \notag \\
& e^{g_{2}\left( \alpha ^{\ast }u_{12}^{\ast }+f_{2}u_{22}^{\ast
}+f_{3}u_{32}^{\ast }\right) +g_{3}\left( \alpha ^{\ast }u_{13}^{\ast
}+f_{2}u_{23}^{\ast }+\allowbreak f_{3}u_{33}^{\ast }\right) }  \notag \\
& e^{\left( \alpha u_{11}+s_{2}u_{21}+\allowbreak s_{3}u_{31}\right)
a_{1}^{\dag }}\left\vert 0\right\rangle _{1}\left\langle 0\right\vert
_{1}e^{\left( \alpha ^{\ast }u_{11}^{\ast }+f_{2}u_{21}^{\ast }+\allowbreak
f_{3}u_{31}^{\ast }\right) a_{1}}  \notag \\
& |_{(s_{2},t_{2},s_{3},t_{3},f_{2},g_{2},f_{3},g_{3})=0}.  \tag{B.3}
\end{align}

\textit{(3) Success probability:}\textbf{\ }Due to \textrm{Tr}$\rho =1$, we
obtain the success probability

\end{subequations}
\begin{subequations}
\begin{align}
p_{n_{2}n_{3},\alpha ,\phi }^{m_{2}m_{3}}& =\frac{e^{-\left\vert \alpha
\right\vert ^{2}}}{m_{3}!m_{2}!n_{2}!n_{3}!}  \notag \\
& \partial _{s_{2}}^{n_{2}}\partial _{s_{3}}^{n_{3}}\partial
_{t_{2}}^{m_{2}}\partial _{t_{3}}^{m_{3}}\partial _{f_{2}}^{n_{2}}\partial
_{f_{3}}^{n_{3}}\partial _{g_{2}}^{m_{2}}\partial _{g_{3}}^{m_{3}}  \notag \\
& e^{t_{2}\left( \alpha u_{12}+s_{2}u_{22}+s_{3}u_{32}\right) +t_{3}\left(
\alpha u_{13}+s_{2}u_{23}+\allowbreak s_{3}u_{33}\right) }  \notag \\
& e^{g_{2}\left( \alpha ^{\ast }u_{12}^{\ast }+f_{2}u_{22}^{\ast
}+f_{3}u_{32}^{\ast }\right) +g_{3}\left( \alpha ^{\ast }u_{13}^{\ast
}+f_{2}u_{23}^{\ast }+\allowbreak f_{3}u_{33}^{\ast }\right) }  \notag \\
& e^{\left( \alpha ^{\ast }u_{11}^{\ast }+f_{2}u_{21}^{\ast }+\allowbreak
f_{3}u_{31}^{\ast }\right) \left( \alpha u_{11}+s_{2}u_{21}+\allowbreak
s_{3}u_{31}\right) }  \notag \\
& |_{(s_{2},t_{2},s_{3},t_{3},f_{2},g_{2},f_{3},g_{3})=0}.  \tag{B.4}
\end{align}

\section*{\textbf{Appendix C: Calculation of }$\langle a^{\dagger
k}a^{l}\rangle _{\protect\rho }$}

In this appendix, we provide two ways to calculate $\langle a^{\dagger
k}a^{l}\rangle _{\rho }$.

\textit{(1) Way 1:}\textbf{\ }Using $\rho $ in Eq.(B.3), we have
\end{subequations}
\begin{subequations}
\begin{align}
\langle a^{\dagger k}a^{l}\rangle _{\rho }& =\frac{e^{-\left\vert \alpha
\right\vert ^{2}}}{m_{3}!m_{2}!n_{2}!n_{3}!p_{n_{2}n_{3},\alpha ,\phi
}^{m_{2}m_{3}}}  \notag \\
& \partial _{s_{2}}^{n_{2}}\partial _{s_{3}}^{n_{3}}\partial
_{t_{2}}^{m_{2}}\partial _{t_{3}}^{m_{3}}\partial _{f_{2}}^{n_{2}}\partial
_{f_{3}}^{n_{3}}\partial _{g_{2}}^{m_{2}}\partial _{g_{3}}^{m_{3}}\partial
_{\mu }^{k}\partial _{\nu }^{l}  \notag \\
& e^{t_{2}\left( \alpha u_{12}+s_{2}u_{22}+s_{3}u_{32}\right) +t_{3}\left(
\alpha u_{13}+s_{2}u_{23}+\allowbreak s_{3}u_{33}\right) }  \notag \\
& e^{g_{2}\left( \alpha ^{\ast }u_{12}^{\ast }+f_{2}u_{22}^{\ast
}+f_{3}u_{32}^{\ast }\right) +g_{3}\left( \alpha ^{\ast }u_{13}^{\ast
}+f_{2}u_{23}^{\ast }+\allowbreak f_{3}u_{33}^{\ast }\right) }  \notag \\
& e^{\left( \nu +f_{2}u_{21}^{\ast }\right) \alpha u_{11}+\left( \allowbreak
\mu +s_{2}u_{21}\right) \alpha ^{\ast }u_{11}^{\ast }+\left( \nu
+f_{3}u_{31}^{\ast }\right) s_{2}u_{21}}  \notag \\
& e^{\left( \mu +s_{3}u_{31}\right) f_{2}u_{21}^{\ast }+\left( \nu +\alpha
^{\ast }u_{11}^{\ast }\right) s_{3}u_{31}+\left( \mu +\allowbreak \alpha
u_{11}\right) f_{3}u_{31}^{\ast }}  \notag \\
& e^{+f_{2}s_{2}\left\vert u_{21}\right\vert ^{2}+\allowbreak
f_{3}s_{3}\left\vert u_{31}\right\vert ^{2}+\left\vert \alpha \right\vert
^{2}\left\vert u_{11}\right\vert ^{2}}  \notag \\
& |_{(f_{2},g_{2},f_{3},g_{3},s_{2},t_{2},s_{3},t_{3},\mu ,\nu )=0}.
\tag{C.1}
\end{align}%
Thus, we can obtain arbitrary $\langle a^{\dagger k}a^{l}\rangle $\ for each
$\left\vert \psi _{i}\right\rangle $ by taking proper $n_{2}$, $n_{3}$, $%
m_{2}$, $m_{3}$.

\textit{(2) Way 2:}\ Using $\rho ^{(h_{l},h_{r})}$, we first have
\end{subequations}
\begin{subequations}
\begin{align}
\left\langle a^{\dag k}a^{l}\right\rangle _{\rho ^{(h_{l},h_{r})}}&
=\partial _{s}^{h_{l}}\partial _{t}^{h_{r}}\partial _{\mu }^{k}\partial
_{\nu }^{l}e^{s\nu +\allowbreak t\mu +st}  \notag \\
& e^{\left( \allowbreak \nu +t\right) \allowbreak u_{11}\alpha +\left( s+\mu
\right) u_{11}^{\ast }\alpha ^{\ast }}|_{(\mu ,\nu ,s,t)=0}.  \tag{C.2}
\end{align}%
and then obtain $\langle a^{\dagger k}a^{l}\rangle _{\rho }$\ according to
Eq.(\ref{2-3}).

\begin{acknowledgments}
This study was supported by the National Natural Science Foundation of China
(No. 11665013).
\end{acknowledgments}

\end{subequations}

\end{document}